\documentclass[prl,twocolumn,nobibnotes]{revtex4}

\begin{document}

\title{Response to ``Comment on
`time-dependent density-matrix renormalization group: a
systematic method for the study of quantum many-body
out-of-equilibrium systems' '' by H. G. Luo, T. Xiang,
and X. Q. Wang}

\author{M. A. Cazalilla}
\affiliation{Donostia International Physics Center (DIPC), 
P. Manuel de Lardizabal 4, 20018 Donostia, Spain}

\author{J. B. Marston}
\affiliation{Department of Physics, Brown University,
Providence, RI 02912-1843, USA}

\maketitle

The authors of the Comment\cite{LXW} propose an improvement to our TdDMRG
algorithm as described in our Letter\cite{Miguel},
providing few details about the proposed extension.
Luo, Xiang, and Wang (LXW) then point out that the oscillations
at late times ($t > 18$) shown in two of the lines plotted in
Fig. 2 of our manuscript are an artifact of the truncation 
of the Hilbert space.

LXW are correct that the oscillations we reported in the one
case of $V/w = 1.1$ are in fact spurious.
We had systematically examined the errors induced by truncations
of the Hilbert space in our Letter.  However we failed to notice that,
for the case of $V/w = 1.1$ when the leads are insulating,
the eigenvalues of the reduced density matrices become so small
so quickly (due to the gap to excitations) that our code
discarded the same portion of the Hilbert space for truncations
M=256 and 512.  Thus in this one case we were left with the illusion 
that we had achieved convergence, when in fact we had not.

Unfortunately the Comment does not provide enough details regarding the
construction of the (reduced) density matrix.
As Eq. 1 of the Comment shows, LXW's density matrix requires knowledge 
of the wavefunction at different times; but to obtain the time-evolved 
wavefunction, one needs the density matrix in the first place. Thus it 
seems that LXW use an iterative procedure:  LXW apparently carry out repeated
integrations forward in time for each chain size starting with
the shortest chain, of length 4 sites.  Once the density
matrix is constructed for a chain of a given length, the chain 
is then lengthened by the addition of two sites via the usual
(infinite-size) DMRG algorithm, and the time evolution is repeated
again from the beginning.  The process is then repeated until a
sufficiently long chain is built up, tailored to the particular
choice of applied bias.  Thus the proposed modification of our TdDMRG
algorithm is computationally intensive and time consuming.  We further point out
that the whole procedure must repeated each time a change is made to
the bias or other parameters, as the density matrix is tailored
to each particular choice of applied bias or parameters. 
This is in contrast to the original TdDMRG procedure
which runs more quickly as it does not
require that the chain be built up again each time the bias or the 
parameters are changed.  And while it seems that LXW have found 
one way to improve the long-time behavior of the TdDMRG algorithm, it is
not at all clear that their procedure is optimal.  It is
unclear, for instance, how much improvement can be expected in
the physical case of electrons with spin, as the Hilbert space
is so much larger in that case.  Other possible
improvements to the original TdDMRG algorithm can be imagined.

Finally, we emphasize that our Letter is substantially correct as 
the Comment corrects only the case $V/w = 1.1$ shown in Fig. 2 of our Letter.
The original TdDMRG scheme itself permits the identification of spurious 
oscillations at late times.  Convergence is tested by checking that 
observables do not change significantly as $M$ is increased.
The results of LXW do not reveal new phenomena not already
presented in our Letter, and it remains to
be demonstrated whether or not the proposed extension is the optimal one.

\end{document}